\documentclass[useAMS,usenatbib]{mn2e}
\usepackage{epsfig,graphicx,lscape}
\title[Fe XVI emission lines in solar spectra]{An investigation of Fe 
XVI emission lines in solar and stellar EUV and soft X-ray spectra}
\author[F. P. Keenan et al.]{F. P. Keenan\thanks{E-mail:
F.Keenan@qub.ac.uk},$^{1}$ J. J. Drake$^{2}$ and 
K. M. Aggarwal$^{1}$
\\
$^{1}$Astrophysics Research Centre, School of Mathematics and Physics, Queen's University, Belfast BT7 1NN
\\
$^{2}$Smithsonian Astrophysical Observatory, MS 3, 60 Garden Street, Cambridge, MA 02138, USA
}

\begin{document}

\date{Accepted 2007 XXXX. Received 2007 XXXX; in original form 2007 XXXX}

\pagerange{\pageref{firstpage}--\pageref{lastpage}} \pubyear{}

\maketitle

\label{firstpage}

\begin{abstract}
New fully relativistic calculations of radiative rates and electron
impact excitation cross sections for Fe\,{\sc xvi} are used to
determine theoretical emission-line ratios applicable to the
251--361\,\AA\ and 32--77\,\AA\ portions of the extreme-ultraviolet
(EUV) and soft X-ray spectral regions, respectively. A comparison of
the EUV results with observations from the Solar Extreme-Ultraviolet
Research Telescope and Spectrograph (SERTS) reveals excellent
agreement between theory and experiment. However, 
for emission lines in the
32--49\,\AA\ portion of the soft X-ray spectral region,
there are large
discrepancies between theory and measurement for both a solar flare spectrum
obtained
with the X-Ray Spectrometer/Spectrograph Telescope (XSST) and 
observations of Capella
from the Low Energy
Transmission Grating Spectrometer (LETGS) on the {\em Chandra X-ray Observatory}.  
These are probably due to
blending in the solar flare and Capella 
data from both first order lines and from
shorter wavelength transitions detected in second and third order. By contrast,
there is very good agreement between our theoretical results and the
XSST and LETGS observations 
in the 50--77\,\AA\ wavelength range, contrary to
previous results.  In particular, there is no evidence that the
Fe\,{\sc xvi} emission from the XSST flare arises from plasma at a
much higher temperature than that expected for Fe\,{\sc xvi} in
ionization equilibrium, as suggested by earlier work.

\end{abstract}

\begin{keywords}
atomic data -- Sun: activity -- Sun: corona -- Sun: ultraviolet -- Sun: X-ray.
\end{keywords}

\section{Introduction}

Emission lines arising from transitions in ions of the sodium
isoelectronic sequence are widely detected in solar ultraviolet,
extreme-ultraviolet (EUV) and soft X-ray spectra (see, for example,
Sandlin et al. 1986; Thomas \& Neupert 1994; Acton et al. 1985). It
has long been known that these lines provide electron temperature
diagnostics for coronal plasmas (Flower \& Nussbaumer 1975), electron
density estimates for laboratory sources (Feldman \& Doschek 1977),
and allow the determination of element abundances (Laming \& Feldman
1999). However, to reliably model these emission features requires
highly accurate atomic data, especially for electron impact excitation
rates (Mason \& Monsignori Fossi 1994).

The most frequently observed Na-like ion in the Sun is Fe\,{\sc xvi},
which is not surprising due to the large Fe cosmic abundance. This ion
provides some of the strongest solar emission features at both EUV and
soft X-ray wavelengths, which have been the topic of several
papers. The most recent examples include the work of Keenan et
al. (2003) on EUV observations from the Solar Extreme-Ultraviolet
Research Telescope and Spectrograph (SERTS), modelled using electron
impact excitation rates calculated with the R-matrix code by Eissner
et al. (1999).  In the soft X-ray spectral region, Cornille et
al. (1997) analysed a solar flare spectrum with excitation rates
determined by these authors in the Distorted-Wave approximation.

However, the atomic physics calculations of Eissner et al. (1999) and
Cornille et al. (1997) for Fe\,{\sc xvi} only considered transitions
among the {\em n} $\leq$ 4 and {\em n} $\leq$ 5 levels,
respectively. More recently, Aggarwal \& Keenan (2006) have generated
excitation rates for Fe\,{\sc xvi} using the fully relativistic Dirac
R-matrix code, which include several improvements over previous
calculations, and in particular the inclusion of all levels with {\em
n} $\leq$ 7. In this paper we use the Aggarwal \& Keenan data to
derive theoretical emission line strengths for Fe\,{\sc xvi}, and
compare these with both solar and stellar EUV and soft X-ray
observations.

\section{Atomic data and theoretical line ratios}

The model ion adopted for Fe~{\sc xvi} consisted of the 39
energetically lowest fine-structure levels arising from the {\em n}
$\leq$ 7 ($\ell$ $\leq$ 4) configurations.  Energies for all these
levels were obtained from the experimental compilation in the NIST
database\footnote{http://physics.nist.gov/PhysRefData/} where
available. However in the case of the 6g $^{2}$G$_{7/2,9/2}$ and 7g
$^{2}$G$_{7/2,9/2}$ levels, energies were taken from the theoretical
results of Aggarwal \& Keenan (2006).  Test calculations including
higher-lying levels were found to have a negligible effect on the
theoretical line ratios considered in this paper.

Electron impact excitation rates for transitions among the levels
discussed above were obtained from Aggarwal \& Keenan (2006), as were
Einstein A-coefficients for allowed and forbidden
transitions. However, we note that the allowed emission lines of
Fe\,{\sc xvi} considered here will be in the coronal approximation
(Elwert 1952), unless the electron density (N$_{e}$) is greater than
10$^{16}$\,cm$^{-3}$ (Feldman \& Doschek 1977), much higher than found
in solar or stellar coronae. Hence the line intensities (and ratios)
will depend primarily on the excitation rates from the ground state,
and will not be sensitive to the adopted A-values.

As has been discussed by, for example, Seaton (1964), proton
excitation may sometimes be important for transitions with small
excitation energie, i.e. fine-structure transitions. However, test
calculations for Fe\,{\sc xvi} setting the proton rates for
$^{2}$P$_{1/2}$--$^{2}$P$_{3/2}$, $^{2}$D$_{3/2}$--$^{2}$D$_{5/2}$,
$^{2}$F$_{5/2}$--$^{2}$F$_{7/2}$ and $^{2}$G$_{7/2}$--$^{2}$G$_{9/2}$
to even 100 times larger than the corresponding electron excitation
rates had a negligible effect on the level populations, showing this
atomic process to be unimportant, at least under solar conditions.

Using the atomic data discussed above in conjunction with a recently
updated version of the statistical equilibrium code of Dufton (1977),
relative Fe\,{\sc xvi} level populations and hence emission-line
strengths were calculated for a range of electron temperatures
(T$_{e}$ = 10$^{6.1}$--10$^{7.1}$\,K in steps of 0.1 dex), over which
Fe\,{\sc xvi} has a fractional abundance in ionization equilibrium of
N(Fe\,{\sc xvi})/N(Fe) $\geq$ 3$\times$10$^{-4}$ (Mazzotta et
al. 1998).  The following assumptions were made in the calculations:
(i) that ionization to and recombination from other ionic levels is
slow compared with bound-bound rates, (ii) that photoexcitation and
induced de-excitation rates are negligible in comparison with the
corresponding collision rates, (iii) that all transitions are
optically thin.

Our results are far too extensive to reproduce here, as with 39
fine-structure levels in our calculations we have intensities for 741
transitions at each of the 11 values of T$_{e}$.  However, results
involving any line pair, in either photon or energy units, are freely
available from one of the authors (FPK) by email on request.  Given
expected errors in the adopted atomic data of at worst $\pm$20 per
cent (see Aggarwal \& Keenan 2006), we would expect our line ratio
calculations to be accurate to better than $\pm$30 per cent.

\section[]{Observational data}

We compare our Fe\,{\sc xvi} line ratio calculations with several data
sets. For EUV transitions we employ observations of several solar
features obtained with the rocket-borne SERTS spectrograph, while for
soft X-ray lines we utilize a solar flare spectrum also taken with
rocket-borne instrument, plus a composite spectrum of Capella from the
{\em Chandra X-ray Observatory}. These data sets are discussed separately
below.

\subsection{Solar EUV observations}

Keenan et al. (2003) have summarised line intensity ratio measurements
involving the Fe\,{\sc xvi} EUV transitions between 251--361\,\AA,
determined from solar spectra of several quiet and active regions, a
small subflare, and an off-limb area, obtained with the SERTS
instrument. The SERTS data sets have spectral resolutions of typically
50--80\,m\AA\ (FWHM) in first-order, and probably provide the best
observations for investigating EUV lines of Fe\,{\sc xvi}. For
example, the {\em Skylab} S082A spectra analysed by Keenan et
al. (1994) had a spectral resolution of only about 200\,m\AA, and the
stronger Fe\,{\sc xvi} emission features were often saturated on the
photographic film used to record the data. The SERTS observations are
of higher spectral resolution and show at worst only a slight
indication of saturation, which does not significant affect the
measured line intensities (Thomas \& Neupert 1994). They have
therefore been adopted in the present analysis. Further information on
the SERTS observations, including details of the wavelength and
absolute flux calibration procedures employed in their reduction, may
be found in Thomas \& Neupert and Brosius et al. (1996, 1998).

In Table 1 we list the Fe\,{\sc xvi} transitions detected in the SERTS
spectra, along with their experimental wavelengths (Thomas \& Neupert
1994), while Table 2 shows the averages of the line ratio values
measured by Keenan et al. (2003). We analyse the averages rather than
the individual ratios for each solar feature detected by SERTS, as the
Fe\,{\sc xvi} emission line intensities are predicted to be
N$_{e}$--insensitive (see Section 2), and could reasonably be expected
to all be formed at the same temperature (that of maximum fractional
abundance in ionization equilibrium). Therefore, the ratio values
should be independent of the solar feature observed. By considering
the average ratios we can reduce the experimental error bars and hence
improve the comparison with theory.

\begin{table}
 \centering
 \begin{minipage}{80mm}
  \caption{Fe\,{\sc xvi} transitions in SERTS spectra.}
  \begin{tabular}{cl}
  \hline
Wavelength (\AA)    &   Transition    
\\
\hline
251.07 & 3p $^{2}$P$_{1/2}$--3d $^{2}$D$_{3/2}$ 
\\
262.98 & 3p $^{2}$P$_{3/2}$--3d $^{2}$D$_{5/2}$ 
\\
265.02 & 3p $^{2}$P$_{3/2}$--3d $^{2}$D$_{3/2}$ 
\\
335.40 & 3s $^{2}$S$_{1/2}$--3p $^{2}$P$_{3/2}$ 
\\
360.75 & 3s $^{2}$S$_{1/2}$--3p $^{2}$P$_{1/2}$ 
\\
\hline
\end{tabular}
\end{minipage} 
\end{table}

\begin{table}
 \centering
\begin{minipage}{80mm}
  \caption{Comparison of theoretical Fe\,{\sc xvi} emission-line intensity
ratios with average values from SERTS.}
  \begin{tabular}{cccc}
  \hline
Line ratio &   Observed & Present & {\sc chianti} 
\\
& & theory$^{a}$ & theory$^{a}$
\\
\hline
335.40/360.75 & 2.2 $\pm$ 0.2 & 2.1 & 2.1
\\
251.07/335.40 & 0.049 $\pm$ 0.008 & 0.049 & 0.044 
\\
262.98/335.40 & 0.066 $\pm$ 0.012 & 0.081 & 0.074
\\
265.02/335.40 & 0.0085 $\pm$ 0.0015 & 0.0078 & 0.0071
\\
\hline
\end{tabular}

$^{a}$Present theoretical ratios and those from {\sc chianti} calculated in energy units
at T$_{e}$ = 10$^{6.4}$\,K and
N$_{e}$ = 10$^{10}$\,cm$^{-3}$.
\end{minipage} 
\end{table}

\subsection{Solar flare soft X-ray observations}

The most detailed measurements to date of Fe\,{\sc xvi} soft X-ray
emission lines remain those made by the X-ray Spectrometer/Spectrograph
Telescope (XSST) during a rocket flight on 1982 July 13 (Acton et
al. 1985).  This instrument recorded the spectrum of an M-class flare
on Kodak 101--07 emulsion, spanning the wavelength range 11--97 \AA\
at a spectral resolution of 0.02 \AA.  The XSST made two exposures,
one of 54~s and the other of 145~s, and the observations presented by
Acton et al.\ are for the latter.  Further details of the XSST
instrument may be found in Brown et al. (1979) and Bruner et
al. (1980), while the reduction and calibration of the observations
are discussed in Acton et al.

In Table 3 we list the Fe\,{\sc xvi} transitions identified in the
XSST spectrum by Acton et al. (1985), along with their measured
wavelengths.  We also indicate possible blending lines or alternative
identifications, as suggested by Acton et al. Additionally, we have
used line lists, such as the NIST database and the Atomic Line List of
Peter van Hoof\footnote{http://www.pa.uky.edu/$\sim$peter/atomic/} to
investigate if any other emission features in the Acton et al. data
set may be due to Fe\,{\sc xvi}. However, only one possible
identification was forthcoming, namely an emission feature at 74.36
\AA\ which is coincident with a predicted Fe\,{\sc xvi} transition.

\begin{table*}
 \centering
\begin{minipage}{140mm}
  \caption{Fe\,{\sc xiii} transitions in the XSST solar flare and {\it
  Chandra} LETGS Capella spectra.}
  \begin{tabular}{cll}
  \hline
Wavelength (\AA)    &   Transition    & Note 
\\
\hline
32.66 & 3p $^{2}$P$_{3/2}$--7d $^{2}$D$_{3/2,5/2}$  & Blend with Ca\,{\sc xii}$^{a}$; 
blend with Fe\,{\sc xvii} and Fe\,{\sc xix} in second-order$^{b}$
\\
34.86 & 3p $^{2}$P$_{1/2}$--6d $^{2}$D$_{3/2}$  &  Possible blend with 
C\,{\sc v}, S\,{\sc xii} and Ar\,{\sc xii}$^{b}$
\\
35.10 & 3p $^{2}$P$_{3/2}$--6d $^{2}$D$_{3/2,5/2}$  & Possible blend
 with Fe\,{\sc xxii} and Fe\,{\sc xxiii} in third-order$^{b}$
\\
35.73 & 3p $^{2}$P$_{1/2}$--6s $^{2}$S$_{1/2}$  & Blend with Ca\,{\sc xi}$^{b}$ 
\\
36.01 & 3p $^{2}$P$_{3/2}$--6s $^{2}$S$_{1/2}$ & Blend with
unidentified feature on blue wing$^{b}$  
\\
36.75 & 3s $^{2}$S$_{1/2}$--5p $^{2}$P$_{3/2}$  & Blend with Fe\,{\sc xvii} in
third-order$^{b}$
\\
36.80 & 3s $^{2}$S$_{1/2}$--5p $^{2}$P$_{1/2}$  & Blend with Fe\,{\sc xvii} in third-order$^{b}$ 
\\
39.83 & 3p $^{2}$P$_{1/2}$--5d $^{2}$D$_{3/2}$  
\\
40.14 & 3p $^{2}$P$_{3/2}$--5d $^{2}$D$_{3/2,5/2}$  &
\\
40.19 & 3d $^{2}$D$_{3/2}$--6f $^{2}$F$_{5/2}$  & Blend with wing
of C\,{\sc v}$^{b}$
\\
40.27 & 3d $^{2}$D$_{5/2}$--6f $^{2}$F$_{5/2,7/2}$ & Blend with C\,{\sc v}$^{a,b}$   
\\
41.90 & 3p $^{2}$P$_{1/2}$--5s $^{2}$S$_{1/2}$ & Blend with Fe\,{\sc xv}$^{a}$; 
possible blend with Fe\,{\sc xviii} and Fe\,{\sc xxi} in third-order$^{b}$
\\
42.27 & 3p $^{2}$P$_{3/2}$--5s $^{2}$S$_{1/2}$  & Blend with Fe\,{\sc xx} and Ni\,{\sc xix} in
third-order$^{b}$
\\
46.66 & 3d $^{2}$D$_{3/2}$--5f $^{2}$F$_{5/2}$ & Blend with wing of 
46.72 \AA\ feature$^{b}$ 
\\
46.72 & 3d $^{2}$D$_{5/2}$--5f $^{2}$F$_{5/2,7/2}$ & Blend with wing of 46.66 \AA\ feature$^{b}$ 
\\
48.97 & 3d $^{2}$D$_{5/2}$--5p $^{2}$P$_{3/2}$ + & Blend with Fe\,{\sc xvii} in third-order$^{b}$      
\\
& 3d $^{2}$D$_{3/2}$--5p $^{2}$P$_{1/2}$ 
\\
50.35 & 3s $^{2}$S$_{1/2}$--4p $^{2}$P$_{3/2}$  & Blend with Fe\,{\sc xvii} in third-order$^{b}$
\\
50.56 & 3s $^{2}$S$_{1/2}$--4p $^{2}$P$_{1/2}$  & Blend with Si\,{\sc x}$^{b}$ 
\\
54.13 & 3p $^{2}$P$_{1/2}$--4d $^{2}$D$_{3/2}$   
\\
54.72 & 3p $^{2}$P$_{3/2}$--4d $^{2}$D$_{5/2}$   
\\
54.77 & 3p $^{2}$P$_{3/2}$--4d $^{2}$D$_{3/2}$   
\\
62.88 & 3p $^{2}$P$_{1/2}$--4s $^{2}$S$_{1/2}$   
\\
63.72 & 3p $^{2}$P$_{3/2}$--4s $^{2}$S$_{1/2}$   
\\
66.25 & 3d $^{2}$D$_{3/2}$--4f $^{2}$F$_{5/2}$   
\\
66.36 & 3d $^{2}$D$_{5/2}$--4f $^{2}$F$_{5/2,7/2}$ & Blend with O\,{\sc vii} in third-order$^{b}$ 
\\
74.36 & 4p $^{2}$P$_{1/2}$--7s $^{2}$S$_{1/2}$ & Blend with N\,{\sc vii} in third-order$^{b}$
\\
76.50 & 3d $^{2}$D$_{5/2}$--4p $^{2}$P$_{3/2}$ & Possible blend with Fe\,{\sc x}$^{b}$ 
\\
76.80 & 3d $^{2}$D$_{3/2}$--4p $^{2}$P$_{1/2}$   & Possible blend with Fe\,{\sc x}$^{b}$  
\\
\hline
\end{tabular}

$^{a}$From Acton et al. (1985) paper on the XSST solar flare spectrum.
\\
$^{b}$From present work on the Capella spectrum.
\end{minipage} 
\end{table*}

The intensity of the 54.72 \AA\ line of Fe\,{\sc xvi} measured by
Acton et al. (1985) in the XSST spectrum is given in Table 4; observed
intensities of the other Fe\,{\sc xvi} transitions may be inferred
from this using the line ratio values provided in the table.  Brown et
al. (1986) note that the relative intensities of lines in the XSST
spectrum similar in strength to the Fe\,{\sc xvi} transitions
discussed here should be accurate to about $\pm$20 per cent, and hence
line ratios to $\pm$ 30 per cent. Evidence for this comes from, for
example, our previous analyses of Ni\,{\sc xviii} and Fe\,{\sc xv}
emission lines in the XSST observations (Keenan et al. 1999, 2006). We
found that, for line ratios involving unblended transitions, agreement
between theory and observation was excellent, with differences that
average only 11 per cent for Ni\,{\sc xviii} and 20 per cent for
Fe\,{\sc xv}.  The observed XSST 
Fe\,{\sc xvi} line ratios in Table 4 have
therefore been assigned a uniform $\pm$30 per cent uncertainty.

\begin{table}
 \centering
\begin{minipage}{80mm}
  \caption{Comparison of theoretical Fe\,{\sc xvi} emission-line intensity
ratios with XSST solar flare and Capella observations.}
  \begin{tabular}{ccccc}
  \hline
Line ratio & Observed & Observed & Present & {\sc chianti} 
\\
 & XSST$^{a}$ & Capella$^{b}$ & theory$^{c}$ & theory$^{c}$
\\
\hline
32.66/54.72 & 0.38 $\pm$ 0.11 & 0.08 $\pm$  0.03 & 0.023 & \ldots
\\
34.86/54.72 & 0.099 $\pm$ 0.030 & 0.09 $\pm$  0.04 & 0.027 & \ldots
\\
35.10/54.72 & 0.17 $\pm$ 0.05 & 0.16 $\pm$  0.04 & 0.055 & \ldots
\\
35.73/54.72 & 0.049 $\pm$ 0.015 & 0.05 $\pm$  0.04 & 0.017 & \ldots
\\
36.01/54.72 & 0.049 $\pm$ 0.015 & 0.11 $\pm$  0.04 & 0.035 & \ldots
\\
36.75/54.72 & 0.22 $\pm$ 0.07 & 0.22 $\pm$  0.04 & 0.089 & 0.094
\\
36.80/54.72 & 0.22 $\pm$ 0.07 & 0.31 $\pm$  0.04 & 0.045 & 0.051
\\
39.83/54.72 & 0.35 $\pm$ 0.11 & 0.12 $\pm$  0.06 & 0.088 & 0.12
\\
40.14/54.72 & 0.55 $\pm$ 0.17 & 0.33 $\pm$  0.07 & 0.18 & 0.24
\\
40.19/54.72 & 0.14 $\pm$ 0.04 & 0.04 $\pm$  0.04 & 0.030 & \ldots
\\
40.27/54.72 & 0.59 $\pm$ 0.18 & 0.93 $\pm$  0.09 & 0.045 & \ldots
\\
41.90/54.72 & 0.14 $\pm$ 0.04 & 0.22 $\pm$  0.21 & 0.077 & 0.16
\\
42.27/54.72 & 0.14 $\pm$ 0.04 & 0.35 $\pm$  0.13 & 0.16 & 0.33
\\
46.66/54.72 & 0.27 $\pm$ 0.08 & 0.28 $\pm$  0.02 & 0.12 & 0.15
\\
46.72/54.72 & 0.37 $\pm$ 0.11 & 0.28 $\pm$  0.03 & 0.17 & 0.23
\\
48.97/54.72 & 0.099 $\pm$ 0.030 & 0.06 $\pm$  0.02 & 0.034 & 0.035
\\
50.35/54.72 & 0.87 $\pm$ 0.26 & 1.35 $\pm$  0.04 & 0.72 & 0.51
\\
50.56/54.72 & 0.44 $\pm$ 0.13 & 0.47 $\pm$  0.03 & 0.38 & 0.28
\\
54.13/54.72 & 0.70 $\pm$ 0.21 & 0.68 $\pm$  0.05 & 0.55 & 0.56
\\
54.77/54.72 & 0.16 $\pm$ 0.05 & 0.07 $\pm$  0.05 & 0.11 & 0.11
\\
62.88/54.72 & 0.88 $\pm$ 0.26 & 0.62 $\pm$  0.07 & 1.1 & 1.6
\\
63.72/54.72 & 1.6 $\pm$ 0.5 & 1.18 $\pm$  0.07 & 2.2 & 3.4
\\
66.25/54.72 & 0.99 $\pm$ 0.30 & 0.73 $\pm$  0.05 & 1.0 & 1.1
\\
66.36/54.72 & 1.3 $\pm$ 0.4 & 1.13 $\pm$  0.05 & 1.6 & 1.7
\\
74.36/54.72 & 0.12 $\pm$ 0.04 & \ldots & 0.0042 & \ldots
\\
76.50/54.72 & 0.36 $\pm$ 0.11 & 0.28 $\pm$  0.04 & 0.26 & 0.17
\\
76.80/54.72 & 0.15 $\pm$ 0.05 & 0.11 $\pm$  0.05 & 0.15 & 0.10
\\
\hline
\end{tabular}

$^{a}$I(54.72 \AA) = 223 $\pm$ 45 photons cm$^{-2}$ s$^{-1}$ arcsec$^{-2}$.
\\
$^{b}$I(54.72 \AA) = $(3.89 \pm 0.17)\times 10^{-5}$ photons
  cm$^{-2}$ s$^{-1}$. 
\\
$^{c}$Present theoretical ratios and those from {\sc chianti} calculated in photon units
at T$_{e}$ = 10$^{6.4}$\,K and
N$_{e}$ = 10$^{10}$\,cm$^{-3}$.
\end{minipage} 
\end{table}

\subsection{Capella spectrum}

The Capella observations analysed here are described in detail by
Keenan et al. (2006).  Spectra were co-added from six separate
observations obtained by the Low Energy Transmission Grating
Spectrograph (LETGS) on the {\em Chandra X-ray Observatory}, employing
the High Resolution Camera spectroscopic detector (HRC-S) in its
standard configuration, yielding a resolution of approximately 0.06
\AA\ (see, for example, Weisskopf et al. 2003 for further details).

The analysis of the co-added 
Capella spectrum followed methods described by Keenan
et al.\ (2006), to which the reader is referred for further details, and
employed the 
PINTofALE\footnote{Available from
http://hea-www.harvard.edu/PINTofALE/} IDL\footnote{Interactive Data
Language, Research Systems Inc.}  software suite (Kashyap \& Drake
2000).  Line fluxes were measured by fitting ``modified Lorentzian'',
or Moffat, functions of the form
$F(\lambda)=a/(1+\frac{\lambda-\lambda_0}{\Gamma})^\beta$, where $a$
is the amplitude, $\Gamma$ a characteristic line width, and
$\beta=2.4$ (Drake 2004).  As in our earlier work, line positions were
allowed to vary from their predicted wavelengths by $\leq$ 0.05 \AA,
this being dictated by imaging distortions in the detector (see, for
example, Chung et al. 2004).  
For lines closely spaced in wavelength, their
relative separations 
were kept fixed to their reference
values, while the position of the group was allowed to vary.

As found in our earlier study (Keenan et al. 2006), lines from the $n=2$ shell of
abundant elements such as Mg, Si, S and Ar plus the $n=3$ shell of Fe,
together with strong features from shorter wavelength transitions arising from
higher spectral orders, are also frequently detected in the 30--80 \AA\
range.  Unlike the {\it Chandra} ACIS CCD detector, the HRC-S
microchannel plate detector possesses no energy resolution of its own
and overlapping spectral orders cannot be separated.  Prior to
performing line fits, we therefore searched for the presence of
significant blends from known strong lines in first and higher orders.
This was undertaken by computing the strengths of lines in the latest version
(Version 5.2) of the {\sc chianti}
database (Dere et al. 1997; Landi et al. 2006)
within our wavelength range
of interest using the differential
emission measure distribution (DEM) of Raassen \& Kaastra
(2007). Relevant blends are noted in Table 3.

As with the XSST observations (Section 3.2), the measured intensity of
the 54.72 \AA\ line is given in Table 4, and those of the other
Fe\,{\sc xvi} transitions may be inferred from this using the line
ratio values.  It should be kept in mind in the interpretation of the
measured fluxes that they are also prone to uncertainty caused by
hidden blends of unidentified lines.  For such cases, there is an
expectation that the measured fluxes might be systematically too high.
Since the resolving power of the {\it Chandra} LETGS is a factor of
approximately 3 lower than that of the XSST, blends tend to be more of a
problem. We discuss this further in Section 4.

\section{Results and discussion}

In Table 2 we list the average observed Fe\,{\sc xvi} EUV
emission-line ratios from the SERTS data sets, along with the
theoretical results at the electron temperature of maximum fractional
abundance for Fe\,{\sc xvi} in ionization equilibrium, T$_{e}$ =
10$^{6.4}$ K (Mazzotta et al. 1998).  Also given in the table are the
calculated ratios from the 
{\sc chianti} database, which
employs the radiative rates and electron impact excitation cross
sections of Sampson et al. (1990). The Sampson et al.\ excitation cross
section results are in very good agreement (generally better than 10
per cent) with those of Cornille et al. (1997), except for 3s
$^{2}$S$_{1/2}$--5s $^{2}$S$_{1/2}$, where there is a difference of 25
per cent at low energies. An inspection of Table 2 reveals excellent
agreement between observation and theory for all of the line
ratios. In some instances there are smaller discrepancies with the
present calculations than with those from {\sc chianti}, and
vice-versa. However, both sets of theoretical results clearly agree with
experiment once the observational uncertainties are taken into
account.

In Table 4 we similarly list the observed Fe\,{\sc xvi} soft X-ray
emission-line ratios from the XSST solar flare and Capella spectra,
along with the present calculations and those from {\sc chianti} at
T$_{e}$ = 10$^{6.4}$ K. We note that the temperature of formation
of the Fe\,{\sc xvi} lines studied here is increased to T$_{e}$ $\simeq$ 10$^{6.7}$ K
when weighted by the Capella DEM, and hence 
theoretical line ratios at this higher value of T$_{e}$ should be employed for comparison
with 
the Capella observations.  However, for the line ratios in Table 4, 
the theoretical values at T$_{e}$ = 10$^{6.4}$ K differ, on average,
by only 12 per cent from those at T$_{e}$ = 10$^{6.7}$ K. For ratios involving
transitions in the wavelength
range 41.90--76.80 \AA, 
the discrepancies are even smaller, averaging only 7 per cent.
As a consequence, changing the theoretical ratios
in Table 4 to values for T$_{e}$ = 10$^{6.7}$ K
would not alter the discussion below.
For example, the theoretical intensity ratio I(32.66 \AA)/I(54.72 \AA)
in Table 4 is 0.023, while the value at T$_{e}$ = 10$^{6.7}$ K is 0.032, 
still much smaller 
than the XSST and Capella measurements of 0.38 and 0.08, respectively. 

The {\sc chianti} results in Table 4 
are not available for all transitions due to
the limited model ion considered by this database, which only incorporates
levels up to and including 5f $^{2}$F$_{5/2,7/2}$.  As noted in
Section 3.2, the 54.72 \AA\ transition has been used as the
denominator for all of the Fe\,{\sc xvi} line ratios, as it is one of
the strongest Fe\,{\sc xvi} emission features in the XSST and Capella
spectral
ranges, and hence should be free from blends.  To confirm this, we have
generated synthetic flare and Capella spectra using {\sc chianti}, which
reveal no nearby emission lines which will contribute more than 3 per
cent to the 54.72 \AA\ intensity.

\begin{figure}
\epsfig{file=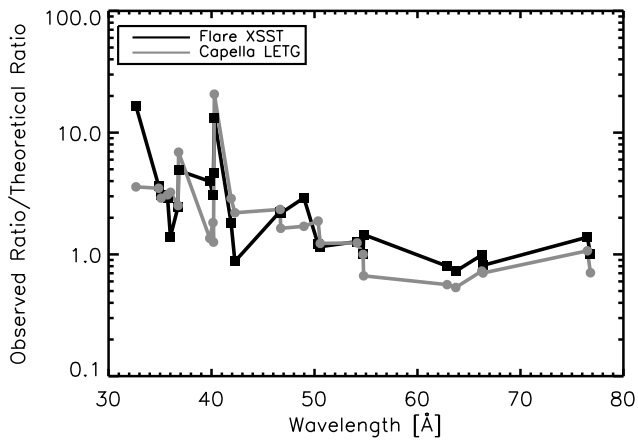,angle=0,width=8.5cm}
\caption{Comparison of observed and predicted ratios of Fe~{\sc xvi}
  line intensities, I($\lambda$)/I(54.72 \AA), for the
XSST spectrum of a
  solar flare and {\it 
  Chandra} observations of Capella, plotted as a function of numerator line
  wavelength. 
}
\end{figure}

An inspection of Table 4 and Fig. 1 (where we plot observed/theoretical 
Fe\,{\sc xvi} line ratios as a function of wavelength)
reveals generally very poor agreement between
theory and observation for line ratios involving Fe\,{\sc xvi}
transitions with wavelengths $\leq$ 48.97 \AA, for both the solar and
Capella spectra.  In the majority of
instances, the measured ratios are larger than the theoretical values
both from the present calculations and from {\sc chianti}.  These
differences are too large to be explained by possible errors in the
adopted electron temperature, as the line ratios are not that
sensitive to T$_{e}$.  
For example, changing T$_{e}$ from 10$^{6.4}$ K
to even as large a value as T$_{e}$ = 
10$^{6.9}$ K only leads to the theoretical intensity ratio I(34.86
\AA)/I(54.72 \AA) increasing from 0.027 to 0.038, still much lower
than the measurements of 0.099 and 0.09 for the Sun
and Capella, respectively.  Similarly, I(36.80 \AA)/I(54.72 \AA)
increases from 0.045 to 0.067, compared to the experimental ratios of 0.22 
for the Sun and 0.31 for Capella.  

One possible cause for the discrepancies is an error in the instrument
intensity calibrations.  Some support for this comes
from discrepancies in several of the Fe\,{\sc xvi} lines when ratioed
against 54.72 \AA, but not when compared with each other.  For
example, the observed values of I(34.86 \AA(/I(54.72 \AA) and I(35.10
\AA)/I(54.72 \AA) are much larger than theory, but the measured
I(34.86 \AA)/I(35.10 \AA) ratios are 0.58 and 0.56 for the XSST and Capella spectra, respectively, 
in good agreement with the
theoretical prediction of 0.49.  On the other hand, I(36.75
\AA)/I(36.80 \AA) involves two lines very close together in
wavelength, yet is predicted to be 2.0 and has measured values of
only 1.0 for the solar flare and 0.71 for Capella.  
Moreover, the {\it Chandra} LETGS calibration is thought to
be accurate to approximately 15\%\ in absolute terms across most of its
band, with relative calibration uncertainties over narrow wavelength
intervals being significantly smaller. It is therefore 
unlikely the larger
discrepancies between theory and observation for Capella can be
explained by calibration errors.

An alternative explanation for the discrepancies is blending of the
Fe\,{\sc xvi} emission features, with instances of good agreement such
as I(34.86 \AA)/I(35.10 \AA) being simply due to coincidence. We
believe this to be the much more likely alternative, as the {\sc chianti}
synthetic solar flare and Capella spectra indicate that most of the
Fe\,{\sc xvi} lines in the 32--49 \AA\ wavelength range are blended
with lines of highly ionized Fe and other ions, such as Fe\,{\sc xvii}
and Fe\,{\sc xxii}, appearing in the observations in
second- or third-order (see Table 3).  Furthermore, Acton et al. (1985) note a
strong second-order spectrum falling between 25--50 \AA\ in the XSST
data set.  In the case of Capella, second-order is suppressed to some
extent by the 1:1 grating bar width-to-spacing ratio, and third-order is
most prominent.  Additionally, it is interesting to note that for the
few Fe\,{\sc xvi} transitions in this wavelength range which the {\sc
chianti} synthetic spectra indicate are not too blended (at least in the XSST observations),
such as
36.01 and 42.27 \AA, agreement between theory and observation is good.

For Fe\,{\sc xvi} transitions with wavelengths $\geq$ 50.35 \AA, an
inspection of Table 4 and Fig. 1 reveals that agreement between the
present theoretical line ratio calculations and the observations
is very good for both the solar and 
Capella spectra, with the
exception of I(74.36 \AA)/I(54.72 \AA).  Our calculations indicate
that the 74.36 \AA\ line is not due to Fe\,{\sc xvi}, with the {\sc
chianti} synthetic flare spectrum suggesting 
an alternative
identification of Ne\,{\sc ix} 1s2s $^{3}$S$_{1}$--1s3p $^{3}$P$_{2}$,
as the predicted intensity ratio of this feature to that of Fe\,{\sc
xv} 73.47 \AA, I(74.36 \AA)/I(73.47 \AA) = 0.41, is in reasonable
agreement with the measured value of I(74.36 \AA)/I(73.47 \AA) = 0.22
$\pm$ 0.07.

For the other line ratios spanning 50.35--76.80 \AA, the discrepancies
between the measurements
and the current calculations average
only 18 per cent for the XSST spectrum and 30 per cent for the Capella observations.
For the latter, the average discrepancy is reduced to 27 per cent if we compare the 
measurements with the theoretical ratios generated at the electron temperature 
appropriate to
the Fe\,{\sc xvi} emission from Capella, i.e. T$_{e}$
= 10$^{6.7}$ K.  
By contrast, there are large differences between
experiment and the {\sc chianti} theoretical results, of 
more than
a factor
of 2 in the case of I(63.72 \AA)/I(54.72
\AA).  The calculations of Cornille et al. (1997) show similar
discrepancies with the XSST measurements, as these authors adopted
similar atomic data to those employed in {\sc chianti} (see Section
2).  Cornille et al. suggested that the Fe\,{\sc xvi} emission in the
XSST flare was at an electron temperature above 10$^{6.7}$ K, as
such a high value of T$_{e}$ was required to provide agreement between
the theoretical and observed line ratios.  However, the present
results show that there is no need to invoke a very high temperature
for the XSST flare, with ratios calculated at the temperature of
Fe\,{\sc xvi} maximum abundance in ionization equilibrium (T$_{e}$ =
10$^{6.4}$ K) being consistent with the observations.

\section{Conclusions}

Theoretical Fe~{\sc xvi} EUV and soft X-ray line intensity ratios based on
new fully-relativistic calculations of radiative rates and electron
impact excitation cross sections have been compared with measurements 
from spectra of the Sun and the binary star
Capella.  Line intensity ratios obtained from SERTS EUV 
observations are in excellent agreement with
the new theoretical values.  In the soft
X-ray regime, lines shortward of 
50~\AA\ in both solar flare and Capella spectra are found to be too
blended by first and higher order transitions, particularly from high ionization stages of
Fe, to provide useful comparisons
with the theoretical ratios.  However, the measured
intensity ratios
I($\lambda$)/I(54.72 \AA) for lines with $\lambda$ $>$ 50 \AA\ are in
good agreement with the new theoretical results, with mean
discrepancies of 18 and 27 per cent
for the flare and Capella observations, respectively.  This 
represents
a considerable improvement over earlier theoretical predictions of the
Fe~{\sc xvi} spectrum.

\section*{Acknowledgments}
KMA acknowledges financial support from EPSRC, while FPK is grateful
to AWE Aldermaston for the award of a William Penney Fellowship.  The
authors thank Peter van Hoof for the use of his Atomic Line List. {\sc
chianti} is a collaborative project involving the Naval Research
Laboratory (USA), Rutherford Appleton Laboratory (UK), Mullard Space
Science Laboratory (UK), and the Universities of Florence (Italy) and
Cambridge (UK), and George Mason University (USA).

\bsp

\label{lastpage}

\end{document}